\documentclass{Rinton-P9x6}
\usepackage{graphicx}
\usepackage{amssymb}

\begin{document}

\title{Casimir force between a sphere and a plane: spectral representation formalism}

\author{Cecilia Noguez and C. E. Rom\'an-Vel\'azquez}
\address{Instituto de F\'{\i}sica, Universidad Nacional Aut\'onoma de M\'exico,\\ Apartado Postal 20-364, M\'exico D.F. 01000,  M\'exico}


\maketitle

\abstracts{
We develop a spectral representation formalism to calculate the Casimir force in the non-retarded limit, between a spherical particle and a substrate, both with arbitrary local dielectric properties. This spectral formalism allows one to do a systematic study of the force as a function of the geometrical variables separately from the dielectric properties. We found that the force does not follow a simple power-law as a function of the separation between the sphere and substrate. As a consequence, the non-retarded Casimir force is enhanced by several orders of magnitude as the sphere approaches the substrate, while at large separations the dipolar term dominates the force.
}

\section{Introduction}
Recent experiments$^{1-4}$ 
to measure the Casimir force have been done choosing a spherical surface and a plate, instead of the plate-plate configuration originally proposed by Casimir\cite{casimir}. The interpretation of the Casimir force in these precision experiments has relied in the Proximity Theorem, developed by Derjaguin and collaborators\cite{proximidad}. The Proximity Theorem estimates the Casimir force per unit area between two curved surfaces of radii $R_1$ and $R_2$, in terms of the Casimir energy per unit area between parallel planes\cite{raul}, ${\cal V}(z)$, where $z$ is the distance between plates. Assuming that the Casimir force on a small area of one curved surface is due to locally ``flat'' portions on the other curved surface, Derjaguin found  that, 
\[F_{\rm PT}(z) = 2 \pi \left(\frac{R_1R_2}{R_1 + R_2}\right) {\cal V}(z).\]
In the limit, when $R_1 =R$, and $R_2 \to \infty$, the problem reduces to the case of a sphere of radius $R$ and a flat plane, that yields to $F_{\rm PT}(z) = 2 \pi R {\cal V}(z)$. The force obtained is a power-law function of the separation, $F_{\rm PT}(z) \propto z^{-\beta}$, where at ``large'' distances  $\beta = 3$ while at short distances $\beta = 2$. This theorem is assumed to hold only when $z \ll R_1, R_2$, however, it is not clear up to what limit this approach is valid. 

The origin of dispersive forces between atoms and macroscopic bodies may be attributed to interactions between the induced charge distributions on them by quantum vacuum fluctuations\cite{casimir}. The charge distribution can be represented, in the simplest approximation, by electric dipoles. This dipole approximation was employed by London \cite{london} to calculate the van der Waals energy $V_{\rm vW} (z)$ between two identical polarizable molecules by using perturbation theory in quantum mechanics. Considering that the molecules interact through a quasi-static potential (or non-retarded limit), London found that $V_{\rm vW}(z) \propto - 1/z^\beta$, where $\beta = 6$  and $z$ is the magnitude of the distance between molecules. In 1948, Casimir and Polder \cite{casypol} studied the interaction of a neutral atom with a perfectly conducting plane, such that, the force is proportional to $ -1/z^\beta$ with $\beta = 4$ in the non-retarded limit \cite{casypol}. Furthermore, they found that the vW interaction could be attributed to the change of the electromagnetic zero-point energy of the classical proper electromagnetic modes of the system. 
The same behavior can be also obtained calculating the change of the zero-point energy between a induced dipole moment on a sphere of polarizability $\alpha$ with its own image dipole in a plane. Contrary to what one may expect, the result for perfect conductors differs from the one obtained using the Proximity Theorem by a factor of $1/z^2$. Therefore, an exact calculation of the Casimir force between a sphere and a planar surface becomes essential. 

In this paper, we study the Casimir force in the non-retarded or quasi-static limit between a sphere and a plane, both with arbitrary dielectric properties. We calculate the difference of the zero-point energy when the two bodies are at a distance $z$, and when they are at infinite, as 
\begin{equation}
{\cal E}(z) = \frac{\hbar}{2}\sum_{s} \left[\omega^{s}(z) - \omega_\infty^s \right], \label{ener}
\end{equation}
where $\omega^s(z)$ and $\omega_\infty^s$,  are the proper electromagnetic modes at $z$ and $z \to \infty$, respectively. By including multipolar interactions to all orders, we calculate exactly the proper electromagnetic modes of the system using a Spectral Representation formalism\cite{ceci2}. We show that this spectral representation formalism has the advantage that separates the contribution of the dielectric properties of the sphere from the contribution of its geometrical properties, allowing to perform a systematic study of the system as a function of the radius of the sphere and its dielectric properties. 

\section{Dipole approximation within the non-retarded limit}

\begin{figure}
\centerline {
\includegraphics[width=3in]{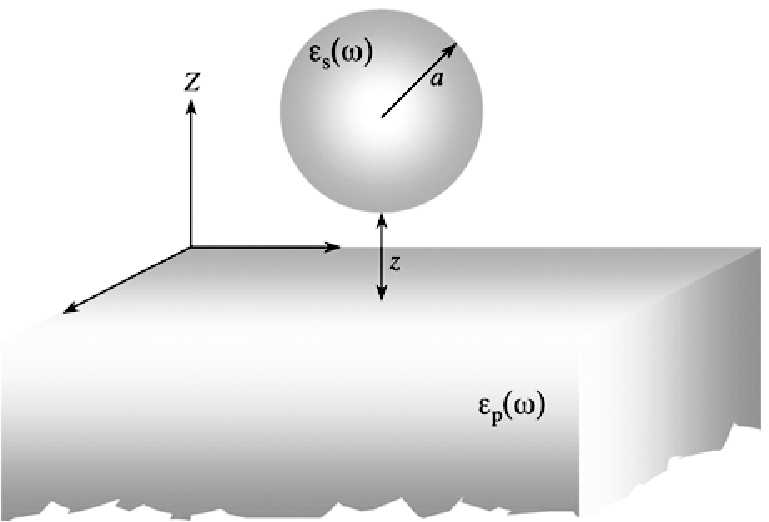}
}
\caption{Schematic model of the sphere-plane system.}
\end{figure}

Let us consider a sphere of radius $a$ and dielectric function $\epsilon_s(\omega)$ which is at a minimum distance $z$ from a semi-infinite plate of dielectric function $\epsilon_p(\omega)$, as shown in Fig.~1. The sphere and plate are electrically neutral. The quantum vacuum fluctuations will induce a charge distribution on the sphere which also induces a charge distribution in the plate, such that, the $lm$-th multipolar moment on the sphere is given by  
\begin{equation}
Q_{lm}  =  \alpha_{lm} \left[ V_{lm}^{\rm vac} + V_{lm}^{\rm sub} \right], \label{qlm}
\end{equation}
where $V_{lm}^{\rm vac}$ is the field associated to the quantum vacuum fluctuations at the zero-point energy, $V_{lm}^{\rm sub}$ is the induced field due to the presence of the plate, and $\alpha_{lm}$ is the $lm$-th polarizability of the sphere. 

In the dipolar approximation, one can find using the image method, that the charge distribution on the plate is described by a dipole moment ${\vec p}_{\rm p}(\omega)$ which is related with the induced dipole moment on the sphere by
\begin{equation}
{\vec p}_{\rm p}(\omega) = \frac{1 - \epsilon_p(\omega)} { 1 + \epsilon_p (\omega)} {\mathbb M} \cdot {\vec p}_{\rm s}(\omega), \label{psub}
\end{equation}
where ${\mathbb M}= (1, 1, -1)$ is a diagonal matrix, and ${\vec p}_{\rm s}$ is the dipole moment on the sphere which, according with Eq.~(\ref{qlm}), is given by
\begin{equation}
{\vec p}_{\rm s}(\omega) =  \alpha(\omega) \left[ {\vec E}^{\rm vac} + {\mathbb T} \cdot {\vec p}_{\rm p}(\omega) \right]. \label{psph}
\end{equation}
Here, $\mathbb T= (3 {\vec r} {\vec r} - r^2{\mathbb I})/r^5$ is a tensor that couples the interaction between the induced dipoles in the sphere and plate, ${\mathbb I}$ is the identity matrix, and ${\vec r}=(0,0,2(z+a))$ is the vector from the center of the sphere to the center of the image charge on the plane, such that, ${\mathbb M}\cdot {\mathbb T}= (-1/r^3, -1/r^3, -2/r^3)$. Substituting Eq.~(\ref{psub}) in Eq.~(\ref{psph}), we find
\begin{equation}
\left[\frac{1}{\alpha} + f_c \frac{b}{r^3} \right]  \mathbb I \cdot {\vec p}_{\rm s} = {\vec E}^{\rm vac}. \label{q2}
\end{equation}
with $f_c = [ 1 - \epsilon_{\rm p}(\omega)]/[  1 + \epsilon_{\rm p}(\omega)]$, and $b= 1$, $1$ or $2$. The eigenfrequencies of the sphere-substrate system are found from the above equation and must be independent of ${\vec E}^{\rm vac}$. The normal frequencies can be obtained when the determinant of the matrix in the left-side of Eq.~(\ref{q2}) is equal to zero. Then, there are three modes (two are degenerated), with frequencies given by the solution of 
\begin{equation}
\left[\frac{1}{{\alpha}(\omega)} +f_c \frac{1}{[2(z+a)]^{3}}\right]^2 
\left[ \frac{1}{{\alpha}(\omega)} +f_c \frac{2}{[2(z+a)]^{3}}\right] = 0. \label{ceros}
\end{equation}
One should notice that the left-side of each term of the above equation  is related only with the dielectric properties of the sphere, while the right-side is related with its geometrical properties only.

\section{Spectral Representation of the sphere-plane configuration}\label{formalism}

The polarizability of a sphere in the non-retarded limit is $\alpha/a^3 \equiv \tilde{\alpha}=[\epsilon_{\rm s} -1]/[\epsilon_{\rm s} + 2]$. Now, let us define the variable $u = [1 - \epsilon_{\rm s}]^{-1}$, then $\tilde{\alpha}(\omega)^{-1} = -3 u(\omega) + 1 $, and substituting in Eq.~(\ref{q2}), one obtains
\begin{equation} 
\left[ - u(\omega) {\mathbb I} + {\mathbb H} \right] \cdot {\bf p}_{\rm s}(\omega) = {\mathbf V}^{\rm vac}, \label{matriz}
\end{equation}
with ${\mathbf V}^{\rm vac} = \frac{1}{3}a^3{\mathbf E}^{\rm vac}$, and ${\mathbb H} = \frac{1}{3} [{\mathbb I} - f_c a^3 {\mathbb T}\cdot{\mathbb M}]$. The later is a dimensionless matrix that only depends on the geometry of the system through the ratio $z/a$. 

Let consider when $f_c$ is real\cite{fc}, then ${\mathbb H}$ is real and symmetric, in such case, we can always find a unitary transformation ${\mathbb U}$ that diagonalizes it,  ${\mathbb U}^{-1}{\mathbb H}{\mathbb U} = n_s$, being $n_s$ the eigenvalues of ${\mathbb H}$. Therefore, the proper modes depend on the separation $z$, and can be obtained from
\begin{equation}
\det{[- u(\omega) {\mathbb I} + \mathbb H]} = \prod_s [-u(\omega) + n_s(z)] =0.
\end{equation}
Alternatively, the eigenfrequencies can be found from the poles of the Green's function of the system defined as $\mathbb G(u) = [ - u(\omega) {\mathbf 1} + {\mathbb H}]^{-1}$, whose $ij$-th element is given by
\begin{equation}
G_{ij}(u) = \sum_s \frac{U_{is}(U_{js})^{-1}}{u-n_s(z)}. \label{green}
\end{equation}
Once the proper modes are obtained, the interaction energy is calculated using Eq.~(1), as the difference between the zero-point energy when the sphere is at a distance $z$ from the substrate and when it is at $z \to \infty$, then the Casimir force is calculated as $F=-d{\cal E}(z)/dz$. 

One should notice that in the spectral representation formalism, the material properties of the sphere are contained in the spectral variable $u$, while the geometrical properties, like the radius of the sphere and the separation of the sphere to the substrate are in ${\mathbb H}$. Furthermore, ${\mathbb H}$ is dimensionless and depends on the ratio $z/R$,  its eigenvalues are independent of ${\mathbf V}^{\rm vac}$. On the other hand, the dielectric properties of the substrate are in $f_c$ which can be a real function even for dispersive materials. Therefore, with the spectral representation one can study separately the contribution of the dielectric properties of the sphere from its geometrical properties, allowing for a systematic investigation of the system.  

\section{Exact calculation within the non-retarded limit}

The $lm$-th multipolar moment induced in the sphere on Eq.~(2) can be written using a multipolar expansion\cite{claro}, that yields to
\begin{equation}
Q_{lm} = \frac{-(2l+1)}{4 \pi} \alpha_{lm} [ V_{lm}^{\rm vac} + \sum_{l', m'} (-1)^{m^{^{\prime }}+l^{^{\prime }}}  A_{lm}^{l'm'} {\hat{Q}}_{l'm'} ],  \label{q}
\end{equation}
where $\hat{Q}_{l'm'}$ is the $l'm'$-th induced multipolar moment in the substrate located at ${\vec r}= (2z+2a,\pi,\varphi)$ from the center of the sphere, and $A_{lm}^{l'm'}$ is the matrix that couples the interaction between sphere and substrate\cite{claro}. The induced $l'm'$-th multipolar moment in the substrate is related with the multipolar charge distribution on the sphere like $\hat{Q}_{l'm'} = (-1)^{l'+m'} f_c Q_{l'm'}$. Substituting this in Eq.~(\ref{q}), one finds 
\begin{equation}
-\sum_{l'm'} \left[\frac{4 \pi \delta_{ll'} \delta_{mm'} }{(2l+1)\alpha_{l'm'}} + f_c A_{lm}^{l'm'} \right] {Q}_{l'm'} = V_{lm}^{\rm vac}.  \label{q3}
\end{equation}
If the sphere is homogeneous its polarizabilities are independent of $m$, and are given by\cite{claro} 
\begin{equation}
\alpha_l (\omega) = \frac{l[\epsilon_{\rm s}(\omega) - 1]} {l [ \epsilon_{\rm s}(\omega) + 1] + 1} a^{2l+1} = \frac{n_{l0}}{n_{l0} - u(\omega)} a^{2l+1}, \label{alfa}
\end{equation}
where $n_{l0} = l/(2l+1)$. The poles of the right-side of Eq.~(\ref{alfa}), $u(\omega_{l0}) = n_{l0}$, yield the frequencies of the proper modes of the isolated sphere. Again, we have separated the material and geometrical properties of the sphere in Eq.~(\ref{alfa}). 

Using Eq.~(\ref{alfa}), we can rewrite Eq.~(\ref{q2}) as 
\[ \sum_{\mu'} [ - u(\omega) \delta_{\mu \mu'}  +  {H}^{\mu '}_{\mu} ] {x}_{\mu'} = b_{\mu}, \] \vskip-.2in
\[{\rm where} \quad \mu \equiv (l,m), \quad x_{\mu} = \frac{Q_{lm}} { (l a^{2l+1})^{1/2}} , \quad b_{\mu} = - \frac{(l a^{2l+1})^{1/2}} {4 \pi} V_{lm}^{\rm vac}, \quad {\rm with} \]
\begin{equation}
H_{\mu}^{\mu'} = n_{l'0}  \delta_{\mu \mu'} + f_c \frac{(a^{l+l'+1})} {4 \pi} (ll')^{1/2} A_{\mu}^{\mu'}. \label{h}
\end{equation}
It was shown in Ref.~12 that, $A_{\mu}^{\mu'}$ is a symmetric matrix that depends on the distance between the center of the ``image-charge'' distribution and the center of the sphere as, $1/[2(z+a)]^{l+l'+1}$. Then, $\mathbb H$ is dimensionless, symmetric, and depends only on the geometry of the system. As we showed, it is possible to find the Green's function, $\mathbb G$, of the system, in terms of  the unitary matrix that diagonalizes $\mathbb H$. The poles of $\mathbb G$,  yield the proper frequencies of the system. 

\section{Results and discussion}

To illustrate the procedure, we use the Drude model for the dielectric function of the sphere, $\epsilon_s(\omega) = 1 - \omega_p^2/[\omega(\omega + i/\tau)]$, with $\omega_p $ the plasma frequency and $\tau$ the relaxation time. In this case, $u(\omega) \omega_p^2 = \omega(\omega + i/\tau )$, such that, the proper modes are given by
\begin{equation}
\omega_s(z) = -i/2\tau + \sqrt{(i/2\tau)^2 + \omega_p^2 n_s(z)} \approx \omega_p \sqrt{n_s(z)},
\end{equation}
with $s = (l,m)$. In the left-side we have considered that $ 1 \gg (\tau\omega_p)^{-1}$. In this case, the fluctuations of the zero-point energy, according with Eq.~(1), are
\begin{equation}
{\cal E}(z) = \frac{\hbar \omega_p}{2} \sum_{l,m}\left[ \sqrt{n_{lm}} - \sqrt{n_{l0}} \right].
\end{equation}
One should notice that it is possible to find the behavior of the energy as a function of $z/a$ for any Drude's sphere independently of its plasma frequency, like $\tilde{{\cal E}} \equiv {\cal E}/\hbar \omega_p $. Furthermore, the Casimir force can be also studied independently of the plasma frequency and radius of the sphere, since ${\cal E}(z/a)$ one can find the force like
\[F=-\frac{d{\cal E}(z/a)}{dz} = -\frac{\partial{\cal E}(z/a)}{\partial(z/a)}  \frac{\partial(z/a)}{\partial z}, \]
and then define the dimensionless variable $\tilde{F}a \equiv F a/ \hbar \omega_p$. In summary, we have shown, for a Drude sphere with $ 1 \gg (\tau\omega_p)^{-1}$, that given the dielectric properties of the substrate the behavior of the energy and force is quite general, showing the potentiality of the Spectral Representation formalism. 

\begin{figure}[htp]
\centerline {
\includegraphics[width=4.8in]{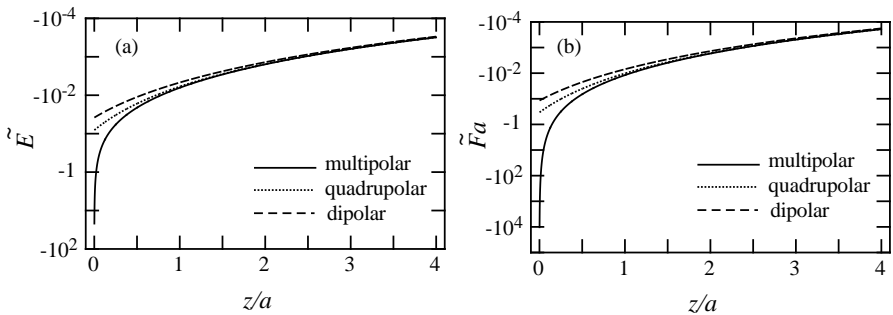}
}
\caption{(a) Energy $\tilde{{\cal E}}$, and (b) force $\tilde{F}a$ as a function of $z/a$.}
\end{figure}

In Fig.~2, we show the energy $\tilde{{\cal E}}$  and force $\tilde{F}a$ for a Drude sphere, of arbitrary plasma frequency and radius over sapphire\cite{sapphire},  as a function of $z/a$. We calculate the energy and force when multipolar interactions up to $l$ and $l' = 2000$ are taken into account, as well as up to dipolar, and up to quadrupolar interactions are considered. When we consider multipolar interactions greater than dipolar ones ($l, l' > 1$), it is observed that $\tilde{{\cal E}}$ and $\tilde{F}a$ do not follow a simple power-law as a function of the separation. We find that $\tilde{{\cal E}}$ and $\tilde{F}a$ are proportional to $1/(z/a)^{\beta}$, where $\beta (z) $ is a positive integer which is also a function of the separation. As the separation decreases ($z \to 0$) the power increases ($\beta \to \infty$). Then, the force can suddenly increases more than four orders of magnitude as compare with the dipolar approximation, when the sphere approaches the substrate. While at large distances ($ 7a > z > 2a$) the force can be obtained exactly if up to quadrupolar interactions are considered. And for $z > 7a$, the interaction between the sphere and the substrate can be modeled using the dipolar approximation only, like in the Casimir and Polder model. Finally, the increment of the force at small separations could explain the physical origin of the large deviations observed in the deflection of atomic beams by metallic surfaces, as well as some instabilities detected in micro and nano devices. However, specific experiments and calculations have to be performed to prove the latter.

\section*{Acknowledgments}
This work has been partly financed by CONACyT grant No.~36651-E and by DGAPA-UNAM grant No.~IN104201.



\begin{thebibliography}{99}
\bibitem{longitud} $a$ and $z$ must be smaller than the characteristic length $l_c$ of the system. For metals, $l_c=c/\omega_p$, where $\omega_p$ is the plasma frequency.
\bibitem{fc} For purely imaginary frequencies $\xi = i \omega$, it is known that $\epsilon(\xi)$ is real, therefore, $f_c$ is also real.
\bibitem{sapphire} The dielectric function of sapphire is real and constant in a wide range of the electromagnetic spectrum, with $\epsilon_s = 3.13$ and $f_c= -0.516$.
\end{thebibliography}
\end{document}